\documentclass{article}
\usepackage{spconf,amsmath,graphicx}
\usepackage{array}
\usepackage{tabu}
\usepackage{amssymb}
\usepackage{url}

\title{3D Vessel Reconstruction in OCT-Angiography via Depth Map Estimation }
\name{Shuai Yu$^{1,5}$,Jianyang Xie$^{1}$,Jinkui Hao$^{1,5}$,Yalin Zheng$^{2}$,Jiong Zhang$^{3}$,Yan Hu$^{4}$,Jiang Liu$^{4}$,Yitian Zhao$^{1*}$}
\address{\small$^1$Cixi Institute of Biomedical Engineering, Ningbo Institute of Materials Technology and Engineering, Chinese Academy of Sciences, \\
\small Ningbo, China *Email: yitian.zhao@nimte.ac.cn\\
\small$^2$Department of Eye and Vision Science, University of Liverpool, Liverpool, UK\\
\small$^3$Keck School of Medicine, University of Southern California, Los Angeles, US\\
\small$^4$Department of Computer Science and Engineering, Southern University of Science and Technology, Shenzhen, China\\
\small$^5$University of Chinese Academy of Sciences, Beijing, China
}

\begin{document}
%
\maketitle
\begin{abstract}
Optical Coherence Tomography Angiography (OCTA) has been increasingly used in the management of eye and systemic diseases in recent years. Manual or automatic analysis of blood vessel in 2D OCTA images (\textit{en face} angiograms) is commonly used in clinical practice, however it may lose rich 3D spatial distribution information of blood vessels or capillaries that are useful for clinical decision-making.
In this paper, we introduce a novel 3D vessel reconstruction framework based on the estimation of vessel depth maps from OCTA images. First, we design a network with structural constraints to predict the depth of blood vessels in OCTA images. In order to promote the accuracy of the predicted depth map at both the overall structure- and pixel- level, we combine MSE and SSIM loss as the training loss function. Finally, the 3D vessel reconstruction is achieved by utilizing the  estimated depth map and 2D vessel segmentation results. Experimental results demonstrate that our method is effective in the depth prediction and 3D vessel reconstruction  for OCTA images.
\end{abstract}
%

\vspace{-0.4cm}
\section{Introduction}
\begin{figure}[htb]
\includegraphics[width=7.5cm]{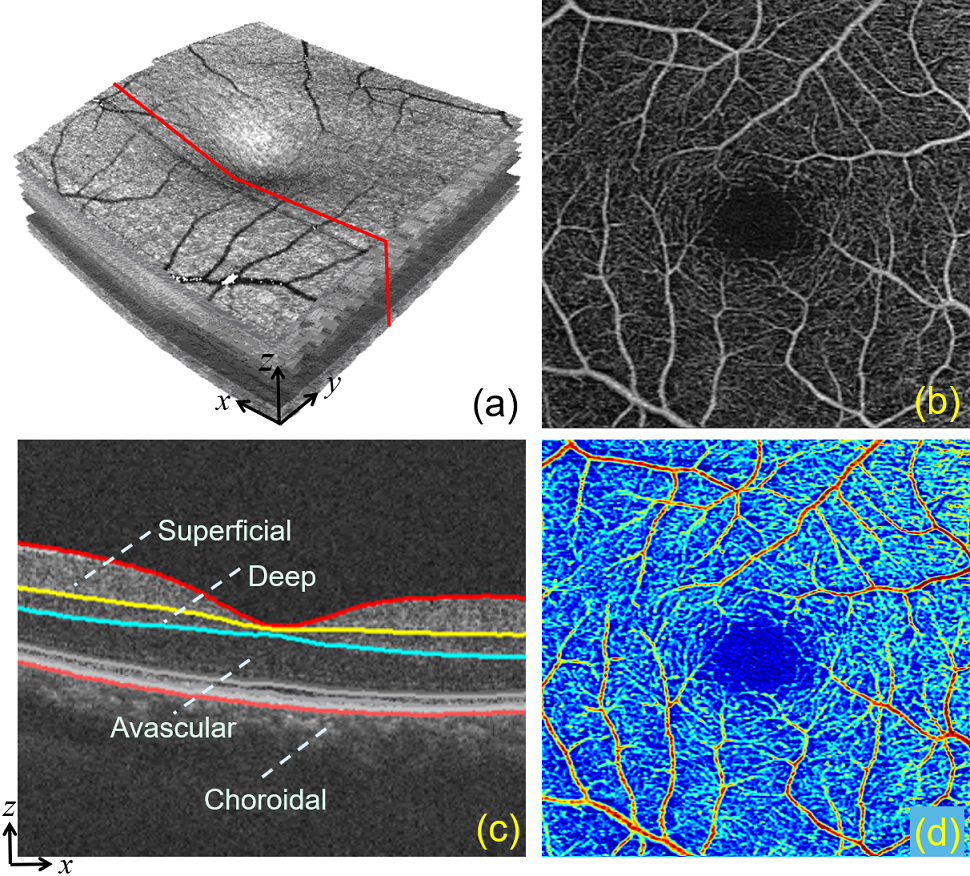}
  \centering
\caption{Visualization of (a) a sample 3D OCTA volume,  (b) 2D \textit{en face} angiogram, (c) different retinal layers, and (d) depth color encoded map of (b).}
\label{fig:OCTA}

\end{figure}

\vspace{-0.2cm}
\label{sec:intro}
The morphological changes of vascular network distributed in the retina is an essential sign to reveal and identify many eye and systemic diseases. Optical Coherence Tomography Angiography (OCTA) is a fast and non-invasive imaging technique that is capable of acquiring blood flow information at capillary-level without injecting contrast agents. In addition, it also produces high-resolution 3D images of the retinal blood vessel down to capillary level, as shown in Fig.\ref{fig:OCTA}(a). Nowadays, OCTA \textit{en face} angiograms, as shown in Fig.\ref{fig:OCTA}(b), has been increasingly used in the studies and clinical diagnosis of various eye-related diseases such as artery and vein occlusions, age related macular degeneration (AMD), diabetic retinopathy (DR), and glaucoma  \cite{de2015review},  \cite{kashani2017optical}.

\begin{figure*}[htb]
\begin{minipage}[b]{1\linewidth}
  \centering
\includegraphics[width=15cm]{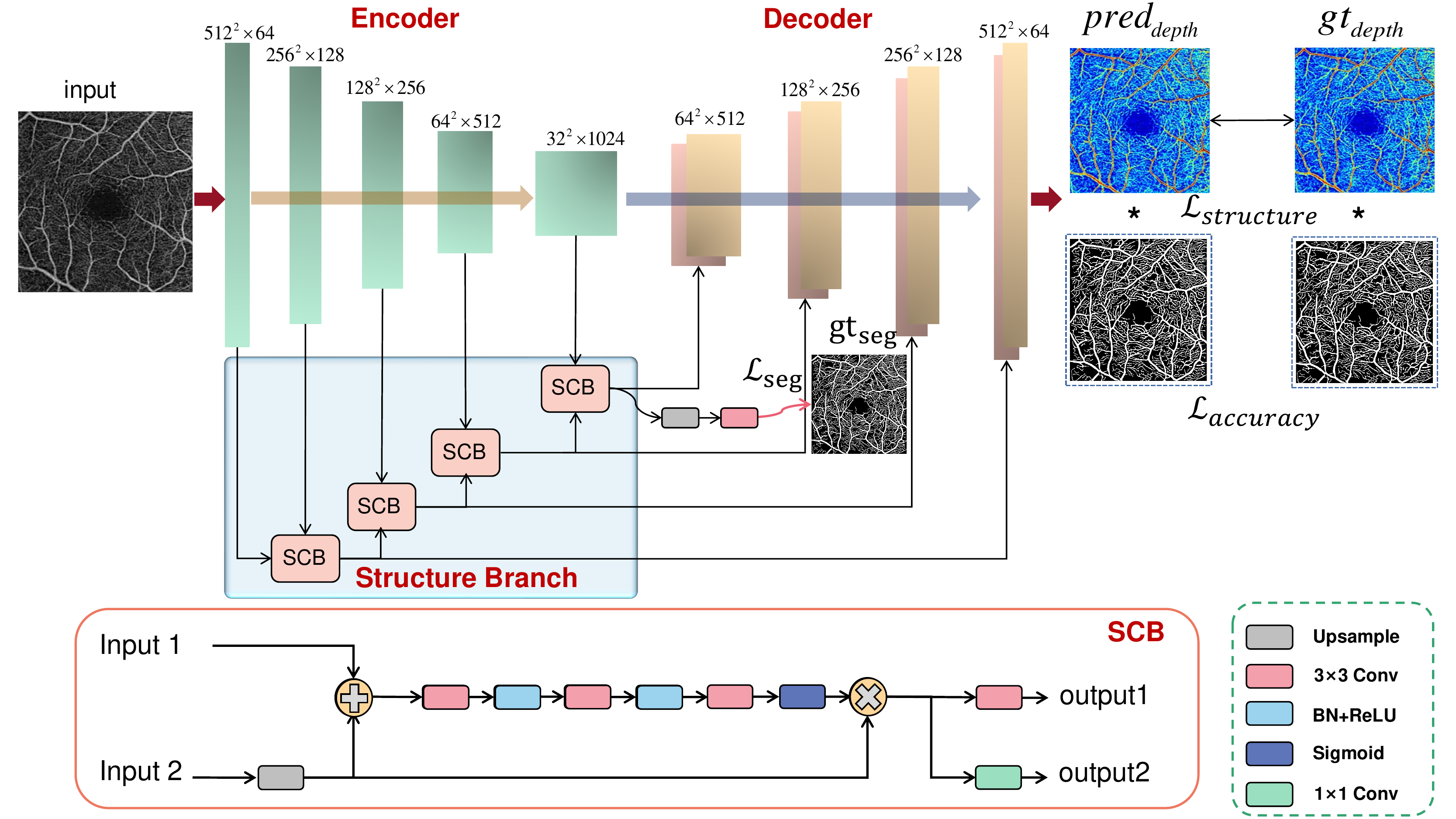}
\end{minipage}
\caption{Overview of the proposed depth estimation method for OCTA image.} 
\label{fig:flowchart}
\end{figure*}

Several studies\cite{kim2016quantifying,zhao2015automated,zhao2019retinal,mou2019cs,xie2020classification,ROSE} have been performed to  analyze retinal vessels in OCTA images in the last decade.
Kim \textit{et al}.\cite{kim2016quantifying} used an intensity-based  algorithm to calculate indices of microvascular density and morphology on \textit{en face} OCTA images. 
Mou \textit{et al}.\cite{mou2019cs} proposed a deep learning-based model to segment the blood vessels on 40 OCTA images. Xie \textit{et al}.\cite{xie2020classification} estimated the vascular topologies of paired color fundus and OCTA images, respectively to classify retinal artery and vein on \textit{en face} OCTA data. Ma \textit{et al}.\cite{ROSE} constructed a first public-available vessel segmentation dataset of retinal OCTA image - ROSE, with aiming to train and validate  automated vessel segmentation algorithms. However, all these works focus on 2D \textit{en face} representation of OCTA only, and they can not recover the 3D spatial information of the vessel. 

3D vessel analysis and visualization may be better at presenting the depth information of vessel abnormalities and can be very helpful for observing microvascular changes\cite{spaide2015volume,zhao2016region,zhao2017automatic}. The rich depth information of blood vessels inevitably be obscured in 2D space because of overlaps during the projection. Therefore, the establishment of 3D OCTA analysis is critical and remains a challenge. In consequence, Zhang \textit{et al}.\cite{zhang20193d} proposed a novel 3D surface-based microvascular segmentation and reconstruction framework, 
with subsequent shape modeling and analysis procedures. However, this study still suffers from shadow projection and may lead to improper vessel structure reconstruction. This is because directly processing of 3D OCTA volume for vessel reconstruction is challenging, due to poor contrast, shadow projection, complex topological structures, and relatively smaller diameters.

By means of OCTA imaging technology, such as the CIRRUS HD-OCT 5000 System (Carl Zeiss Meditec), equipped with AngioPlex® OCT Angiography software, a depth encoded color map is provided (we refer it as depth map in this paper), as illustrated in Fig.\ref{fig:OCTA} (d). Depth map is a color encoded slab with different colors representing different depth layers: superficial retina, deep retina, avascular retina, choriocapillaris, and choroid layer, and these layers can be seen in Fig.\ref{fig:OCTA} (c). Similar to the range image in computer vision, depth image refers to an image whose pixel value is the distance from the imaging sensor to each position in the scene. For depth map obtained by CIRRUS HD-OCT 5000 System, color red indicates the vessels are closer to the imaging sensor and blue represents the vessels that are further.


In this work, we aim to estimate 3D vessel network  of each OCTA \textit{en face} angiogram by the prediction of depth map.  
To the best of our knowledge, this is the first work to obtain vessel depth map and to generate  3D vessel structure from OCTA \textit{en face} angiogram.  We introduce a deep neural network with structural constraints for predicting the depth map of \textit{en face} angiogram, and a novel combined loss is utilized to train the network to ensure that the depth map of the blood vessel is close to its groundtruth (generated by CIRRUS HD-OCT 5000 System). The spatial position of the vessel in 3D space is obtained from the depth image, and the 3D vessel structures are finally reconstructed by ensembling 2D vessel segmentation results and the predicted depth maps.

\vspace{-0.4cm}
\section{Method}
\label{sec:method}
\vspace{-0.2cm}
\subsection{Depth Map Estimation Network}
\vspace{-0.2cm}
\label{ssec:network}

In this section, we present our structure-constraint CNN architecture for depth map estimation. As illustrated in Fig.\ref{fig:flowchart}, our network consists of three components: feature encoder module, feature decoder module, and the structure constraint blocks (SCB). We further design a specific loss function for training the model to promote the predicted depth map accuracy at both the overall structure- and pixel- level. 
\vspace{-0.3cm}
\subsubsection{Network architecture}
\vspace{-0.1cm}
In view of U-Net's\cite{ronneberger2015u} outstanding performance in processing medical images, we employ the same architecture configuration with it as the encoder and decoder. 
Since retinal blood vessels form a very complicated topological structure, the accuracy of depth prediction for the part of the image with blood vessels is particularly important and challenging.
In order to predict the depth of each pixel located in vessel structure, we further utilize a structure branch to process blood vessel information in the form of semantic structures. We enforce structure branch to only process vessel-related information by our carefully designed SCB and local supervision.

Different from skip connection used in\cite{ronneberger2015u}, SCB enables the decoder to process relevant information only, and we use SCB after every block of the encoder. Let $e_t$ $(t\in {2,..., 5})$ denote the output of $t_{th}$ encoder block, and $s_{\widetilde{t}}$ denote the corresponding intermediate representations of structure branch. We first obtain an attention map $a_{t-1}$ by concatenating $e_t$ and $s_{\widetilde{t}}$ followed by convolutional layers, batch normalization and nonlinear activation layers. Given the attention map $a_{t-1}$, an element-wise product is applied between $e_t$ and $a_{t-1}$ to acquire weighted map. Note that upsampling is employed on $e_t$ before concatenation, to ensure $e_t$ and $s_{\widetilde{t}}$ has the same size. Since containing rich edge information, the low-level features from the first block of the encoder are used to obtain the initial weighted map. Intuitively, $a_{t-1}$ can be seen as an attention map that weights more important areas with blood vessel information. The filtered feature maps by SCB(i.e., output2 in Fig.\ref{fig:flowchart}) are cascaded with the corresponding decoder feature maps to provide refined structure-related information.  
And the output of the last SCB is subjected to upsampling and convolution operation to obtain the blood vessel prediction map.


\vspace{-0.3cm}
\subsubsection{Loss function}
\label{sssec:loss}
In this work, we jointly supervise depth map estimation and vessel segmentation during training.  We use mean square error (MSE) loss on predicted blood vessel maps $pred_{seg}$:
\begin{equation}\small
\mathit{\mathcal{L}_{seg} =\mathcal{L}_{MSE}(pred_{seg},gt_{seg})},
\end{equation}
where $gt_{seg} \in \mathbb {{R}^{H \times W}}$ denotes groundtruth of blood vessel, and the generation of groundtruth will be introduced in Section 3.1. To promote the accuracy of the predicted depth maps, we constrain them at both the pixel- and overall structure- level. Specifically, MSE loss is utilized to ensure the pixel level accuracy of depth map:
\begin{equation}\small
\mathit{\mathcal{L}_{accuracy}=\lambda_{1}\mathcal{L}_{MSE}(v,\widetilde{v})+\lambda_{2}\mathcal{L}_{MSE}(b,\widetilde{b})},
\end{equation}
where $v$ and $b$ represent predicted depth map of vessel and background areas, which are generated by multiplying predicted masks of structure branch, and $\widetilde{v}$ and $\widetilde{b}$ are corresponding groundtruth. We aim to ensure the model pay more attention to the vessels than the background, and different weights $\lambda_{1}$ and $\lambda_{2}$ are utilized in $\mathcal{L}_{accuracy}$. In our experiments, $\lambda_{1}$ and $\lambda_{2}$ are empirically set to 0.8 and 0.2, respectively.


Since the MSE has difficulty in discriminating structural content in images, we add structural similarity index measure (SSIM) \cite{ssim} loss to ensure the accuracy of the predicted images
in an overall structure manner. SSIM compares two images from the perspectives of brightness, contrast, and structure. 
Therefore, we define the structural loss between the predicted depth map $pred_{depth}$ and the ground truth $gt_{depth}$ as:
\begin{equation}\small
\mathit{\mathcal{L}_{structure} =SSIM(pred_{depth},gt_{depth})},
\end{equation}
and the final loss function may be defined as:
\begin{equation}\small
\mathit{\mathcal{L}_{total} =\mathcal{L}_{seg}+\mathcal{L}_{accuracy}+\mathcal{L}_{structure}}
\end{equation}

\vspace{-0.3cm}
\subsection{3D Vessel Reconstruction}
\vspace{-0.2cm}
\label{sssec:3D_reconstruction}

The 3D vessel reconstruction process may be treat as a mapping problem from the segmented 2D vessel to 3D space.
As we aforementioned, our network is also able to detect the vessel structures, and we use skeletonization method \cite{Peter2012Fast} to extract centerline of vessels in OCTA \textit{en face} images, and the bifurcation points of vessel network can be extracted by locating intersection points (pixels with more than two neighbours). All the intersection points and their neighbours may then be removed from the centreline map, in order to obtain an image with clearly separated vessel segments. Therefore, 3D point clouds composed of the centreline points may be obtained by utilizing the predicted depth map, where adjacent segments were linked according to the topology consistency, and the \textit{Tubefilter} process in VTK \footnote{\url{https://vtk.org/}} software is applied to rend the 3D OCTA vessels architecture. In each vessel segment, the bilinear interpolation was utilized to ensure vessel continuity. 

\begin{figure*}[htb]
\begin{minipage}[b]{1\linewidth}
  \centering
\includegraphics[width=14cm]{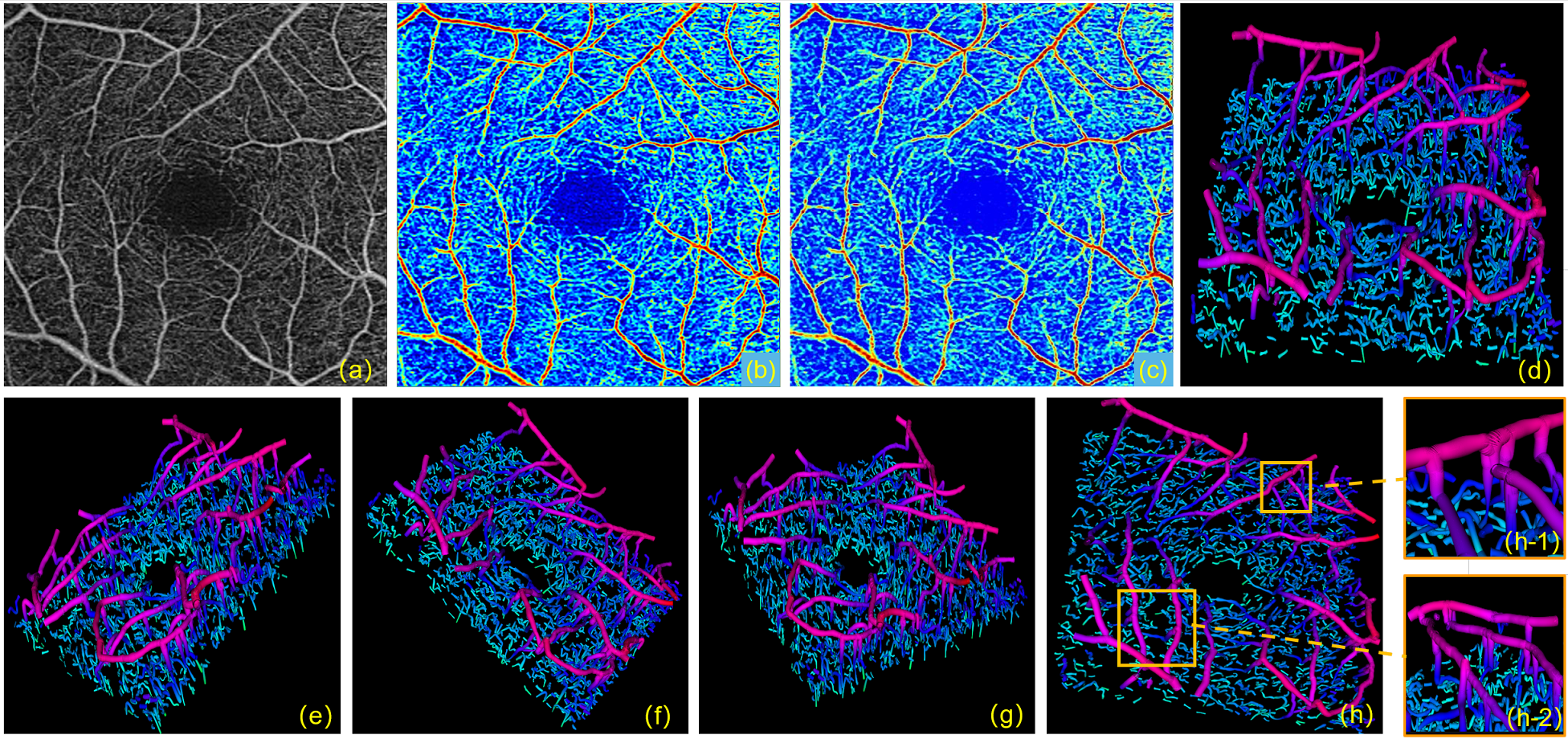}
\end{minipage}
\caption{Illustration of depth map and 3D vessel reconstruction. (a) \textit{en face} OCTA angiogram. (b) Groundtruth of depth map. (c) Predicted depth map by our method. (d)-(h) Reconstructed 3D vessels with different angle of view.}
\label{fig:Experimental_diagram}
\end{figure*}

\begin{table*}[t]\small
\centering
\setlength{\tabcolsep}{0mm}{
\begin{tabular}{|c||l|l|l|l|l|l|}
\hline
              & Eigen et al. \cite{eigen2014depth}      & {Eigen et al.} \cite{eigen2015predicting}      &{ Laina et al.} \cite{laina2016deeper}       &{Chen et al.} \cite{chen2016single}       &{U-Net} \cite{ronneberger2015u}       &{Our method}       \\ \hline\hline
\textbf{ACC} ($\delta_1$ $<$ 1.25 ) & \qquad 0.441 & \qquad 0.596 & \qquad 0.925 & \qquad\textbf{0.973} & \quad 0.918 & \quad 0.971 \\ 
\hline
\textbf{ACC} ($\delta_2$ $<$ $1.25^{2}$) & \qquad0.783 & \qquad0.854 &\qquad 0.967 & \qquad0.985 &\quad 0.987 & \quad\textbf{0.993} \\ 
\hline
\textbf{ACC} ($\delta_3$ $<$ $1.25^{3}$) & \qquad0.927 &\qquad 0.941 &\qquad 0.981 & \qquad0.991 &\quad0.992 & \quad\textbf{0.996} \\  \hline
\textbf{ARD} & \qquad0.392 &\qquad 0.310 &\qquad0.138 & \qquad0.114 & \quad0.091 & \quad\textbf{0.058} \\ \hline
\textbf{RMSE}    & \qquad0.378 &\qquad 0.343 &\qquad 0.213 &\qquad 0.138 & \quad0.152 &\quad\textbf{0.107} \\ \hline\hline
\textbf{CD} &\qquad2.988 &\qquad2.872       &\qquad2.532       &\qquad 1.302       &\quad1.501       &\quad\textbf{1.192}       \\  \hline
\textbf{HD}&\qquad5.273       &\qquad4.024       &\qquad4.571       &\qquad4.309       &\quad3.430       &\quad\textbf{3.207}       \\ \hline
\end{tabular}}
\caption{ Depth map estimation and 3D vessel reconstruction results by different methods.}
\label{table:results}
\end{table*}

\vspace{-0.3cm}
\section{Experiments}
\label{sec:Experiment}

\vspace{-0.2cm}
\subsection{Dataset and Evaluation Metrics}
\vspace{-0.2cm}
\label{sssec:datasets}
A dataset comprising of 80 pairs of OCTA \textit{en face} angiograms and their depth maps were used in this work. All the OCTA and depth maps were obtained by the CIRRUS HD-OCT 5000 System (Carl Zeiss Meditec Inc., USA), equipped with AngioPlex® OCT Angiography, with an image resolution of $512 \times 512$ pixels, and the scan area was $3\times3$ mm centered on the fovea.
We divide these 80 pairs images into training and testing sets, 56 of which were used for training, and the rest for testing. It is worth noting that a state-of-the-art OCTA vessel segmentation model OCTA-Net\cite{ROSE} was used to extract vessels of training set, and an image analysis expert further refined the vessel segmentation results as groundtruth. 

Seven metrics were employed to validate the performance of the proposed framework, five of which were used for the evaluation of depth prediction and two were used for the validation of 3D vessel reconstruction.
The \textbf{accuracy} (ACC) metric $\delta$\cite{ladicky2014pulling} was employed to validate the proposed depth prediction method:  $\delta=max(\frac{D_{i}}{D_{i}^*},\frac{D_{i}^*}{D_{i}})<T$, where $D_{i}$ and ${D_{i}^*}$ are the estimated depth and the corresponding depth of the \textit{i}-th pixel of the groundtruth, respectively. As suggested in\cite{laina2016deeper}, three different thresholds $T$ (1.25, $1.25^{2}$, $1.25^{3}$) were used in the accuracy metric.  As the most commonly-used metrics in evaluating monocular image depth estimation, the \textbf{Absolute Relative Difference} (ARD) and \textbf{Root Mean Squared Error} (RMSE) are also  used in this work. In order to evaluate the 3D vessel reconstruction, \textbf{Chamfer Distance} (CD)\cite{borgefors1986distance} and \textbf{Hausdorff Distance} (HD)\cite{huttenlocher1993comparing} were further employed. CD and HD are both metrics to describe the similarity between two sets of points. 

\subsection{Experimental Results}
\vspace{-0.2cm}
\label{sssec:Experimental_results}

Fig.\ref{fig:Experimental_diagram} illustrates the proposed framework in predicting depth map and 3D vessel reconstruction. It may be seen that our method is able to generate a depth map similarly to the groundtruth depth map, as demonstrated in Fig.\ref{fig:Experimental_diagram} (b) and (c). However, it is difficult to demonstrate conclusively the superiority of the proposed depth estimation method purely by the above visual inspection, we have compared our depth prediction results with those produced by other four state-of-the-art
approaches which were proposed by Eigen \textit{et al}.\cite{eigen2014depth}, Eigen \textit{et al}.\cite{eigen2015predicting}, Laina \textit{et al}.\cite{laina2016deeper} and Chen \textit{et al}.\cite{chen2016single}. In addition, as the backbone of our work, U-Net\cite{ronneberger2015u} was also employed as one of the benchmark approaches. Table \ref{table:results} reports the evaluation results in terms of five different depth prediction metrics. As can be observed, the proposed method has reached the best performance in terms of all the metrics by significant margins, with only a single exception: its ACC ($\delta_1$ $<$ 1.25) score is 0.002 lower than that of Chen \textit{et al}.\cite{chen2016single}.  

Fig.\ref{fig:Experimental_diagram}(d)-(h) demonstrate the 3D vessel reconstruction results with different angle of views. Our method is able to produce high vessel visibility on both large vessels and small capillaries. As can be observed in Fig.\ref{fig:Experimental_diagram}(h-1) and (h-2), the enlarged representative regions show high preservation of different scales of vessel and bifurcation structure. The distributions of large and small vessels are at different depth gradually decrease from the parafovea to the fovea, and this is in line with the anatomy of the retina.
The CD and HD metrics also reveal that our method is superior in 3D vessel reconstruction  when compared with other methods. The structural constraints and combined loss of our method substantially play very important role  in depth prediction.


\vspace{-0.3cm}
\section{Conclusion}
\vspace{-0.3cm}
\label{sec:Conclusion}
In this work, we have proposed a novel framework to reconstruct 3D vessel structure in OCTA via a depth prediction network. The significance
of our method is that this work may be considered as the first attempt to predict the vessel depth information on 2D \textit{en face} angiograms.
The uncertainty of the position of blood vessels in 3D spatial domain is estimated by combining 2D vessel segmentation and the predicted depth map. The
high evaluation performance in terms of depth map prediction and 3D vessel reconstruction demonstrate the effectiveness of our method. It shows the
great potential of exploring 3D vessel analysis in clinical practice, and we will focus on using the proposed framework for the diagnosis of eye-related disease in clinical settings.

\section{Compliance with Ethical Standards}
This study was performed in line with the principles of the Declaration of Helsinki. Approval was granted by the Ethics Committee of Ningbo Institute of Materials Technology and Engineering, Chinese Academy of Sciences.
\section{Acknowledgments}
This work was supported in part by Zhejiang Provincial Natural Science Foundation of China under Grant (LZ19F010001, LQ20F030002, LQ19H180001), in part by  Ningbo Public Welfare Science and Technology Project (2019C50049), and in part by the Ningbo 2025 S\&T Megaprojects under Grant (2019B10033, 2019B10061), and in part
by the Ningbo Natural Science Foundation under Grant 2019A610354.


\bibliographystyle{ieeetr}
\bibliography{refs}
\end{document}